\documentclass[12pt]{article}
\usepackage{amssymb}
\usepackage{amsmath}
\usepackage{hyperref}
\usepackage{graphicx}
\usepackage{placeins}
\usepackage{bm}
\usepackage{latexsym}
\begin{document}
\title{\bf  Hydrodynamics of Arcsin AdS Black Brane }

\author{Mehdi Sadeghi\thanks{Corresponding author: Email: mehdi.sadeghi@abru.ac.ir  }\\
	{}\\
	{\small {\em  Department of Physics, Faculty of Basic Sciences,}}\\
	{\small {\em  Ayatollah Boroujerdi University, Boroujerd, Iran}}
}
\date{\today}
\maketitle

\abstract{In this paper, we explore a modified black brane within AdS spacetime, characterized by the Lagrangian density $\frac{1}{q} \text{arcsin}(qR)-2\Lambda$. Due to the absence of an analytic solution, we approach the Einstein equations using a perturbative method, extending our analysis to the second order in $q$. Subsequently, we compute the ratio of shear viscosity to entropy density. Our results suggest that the KSS Bound is not saturated in this model.}\\



\section{Introduction} \label{intro}

In the quest to understand the fundamental nature of gravity and its interplay with quantum field theories, modified theories of gravity have gained significant attention. Among these, arcsin-gravity \cite{Kruglov:2014rja} emerges as a compelling alternative. This modified gravity \cite{Rastall:1972swe},\cite{Lovelock:1971yv},\cite{Stelle:1976gc},\cite{Hu:2007nk},\cite{Brans:1961sx},\cite{Sadeghi:2023tzf} theory is characterized by a distinctive non-linear structure that differs from classical general relativity, as it introduces higher-order terms that can account for phenomena often attributed to dark energy and dark matter. The motivation for exploring arcsin-gravity springs from the need to investigate potential footprints of new physics in gravitational interactions, especially in the context of holographic principles.\\
One of the profound insights arising from theoretical research in gravity is the AdS/CFT duality\cite{Maldacena:1997re}, proposed by Juan Maldacena. This duality posits a remarkable correspondence between certain string theories formulated in Anti-de Sitter (AdS) space and conformal field theories (CFT) defined on the boundary of this space. This relationship provides a powerful framework through which researchers can transfer problems from gauge theories, considered to be difficult to solve, to gravitational theories, where the geometry provides a clearer understanding of dynamics. In particular, the duality allows us to leverage the geometric nature of gravity to address issues in particle physics, including twinning aspects of strong coupling and quantum field theory.\\
One significant application of the AdS/CFT correspondence is the study of transport coefficients, among which shear viscosity plays a crucial role. Shear viscosity is a measure of a fluid's resistance to shear flow, and its behavior in the context of quantum field theories can illuminate vital features of the underlying interactions. The Kovtun-Son-Starinets (KSS) bound, which states that the ratio of shear viscosity to entropy density $\frac{\eta }{s}$ must satisfy $\frac{\eta }{s}=\frac{1 }{4 \pi}$ \footnote{We consider $\hbar=k_B=1$.}, is particularly noteworthy as it highlights the fundamental connection between thermodynamic properties and quantum aspects of the theory. The significance of this ratio extends beyond purely theoretical realms, as it has implications for the behavior of strongly coupled plasmas, such as those created in heavy-ion collision experiments.\\
Observational data regarding the ratio of shear viscosity to entropy density ($\frac{\eta }{s}=\frac{1 }{4 \pi}$) have been derived from experimental findings, particularly in the context of strongly coupled plasmas, such as those produced in heavy-ion collisions at facilities like the Large Hadron Collider (LHC) and the Relativistic Heavy Ion Collider (RHIC). These experiments suggest that the quark-gluon plasma (QGP), a state of matter formed under extreme temperature and density, exhibits a remarkably low ratio of shear viscosity to entropy density, approaching the KSS bound limit of $\frac{1 }{4 \pi}$\cite{Kovtun:2004de}\cite{Policastro2002}. This behavior indicates that the QGP behaves almost like a perfect fluid, with minimal resistance to flow, showcasing near-ideal characteristics. Recent measurements have yielded values that are consistent with theoretical predictions emerging from holographic principles, supporting the notion that the QGP behaves according to quantum field theory principles. Furthermore, the results underscore the crucial connections between fundamental quantum properties and macroscopic fluid dynamics, providing valuable insights into the underlying mechanisms of particle interaction and the dynamics of the early universe. Such findings not only confirm the theoretical frameworks but also pave the way for future investigations into non-ideal fluids in quantum contexts.\\
In our work, we investigate the arcsin AdS black brane solution to derive the ratio of shear viscosity to entropy density for our modified gravity model. By exploring its implications in the context of the KSS bound, we aim to understand how the non-linear characteristics of arcsin-gravity influence transport properties. Furthermore, examining when and how the KSS bound is saturated or violated offers insights into the coupling dynamics of the dual field theory. Our analysis not only tests the robustness of the AdS/CFT correspondence in the context of modified gravity but also sheds light on the broader implications of such gravitational theories in understanding the fabric of spacetime and its interaction with quantum fields. Through this investigation, we aspire to contribute to the growing body of literature that seeks to bridge the gap between classical gravitational models and their quantum counterparts, probing the frontiers of fundamental physics.\\
In summary, the exploration of arcsin-gravity in the context of the AdS/CFT duality presents a unique opportunity to glean insights into the dynamics of modified theories of gravity, the role of shear viscosity in quantum field theories, and the fundamental principles governing the Universe.
ty, quantum field theory, and thermal transport in strongly correlated systems.\\
\section{ $\frac{1}{q}\arcsin(qR)$-AdS Black Hole}
\label{sec2}

\indent The four-dimensional action representing arcsin-gravity as a modified gravity with a negative cosmological constant is given by \cite{Kruglov:2014rja},
\begin{eqnarray}\label{action}
S=\frac{1}{16\pi G}\int d^{4}  x\sqrt{-g} \bigg[\frac{1}{q} \arcsin(qR)-2\Lambda\bigg],
\end{eqnarray}
where $R$ denotes the Ricci scalar, $\Lambda=-\frac{3}{L^2}$ is the cosmological constant, and $L$ represents the AdS radius. As $q \to 0 $, the exponential term reduces to the standard Einstein-Hilbert gravity.\\
The equations of motion are derived by performing a variation of the action \ref{action} with respect to $g_{\alpha \beta }$,
\begin{align}\label{EH}
	& \frac{R_{\alpha \beta }}{(1 -  q^2 R^2)^{1/2}}+\Lambda g_{\alpha \beta} +\frac{1}{q} \arcsin(q R) g_{\alpha \beta} -  \frac{3 q^4 R^2 \nabla_{\alpha }R \nabla_{\beta }R}{(1 -  q^2 R^2)^{5/2}}  \nonumber \\&-  \frac{q^2 \nabla_{\alpha }R \nabla_{\beta }R}{(1 -  q^2 R^2)^{3/2}} -\frac{q^2 R \nabla_{\beta }\nabla_{\alpha }R}{(1 -  q^2 R^2)^{3/2}} + \frac{3 q^4 g_{\alpha \beta } \
		g^{\gamma \lambda } R^2 \nabla_{\gamma }R \nabla_{\lambda }R}{(1 -  q^2 R^2)^{5/2}}  \nonumber \\& + \
	\frac{q^2 g_{\alpha \beta } g^{\gamma \lambda } \nabla_{\gamma }R \nabla_{\lambda }R}{(1 -  q^2 R^2)^{3/2}}+ \frac{q^2 g_{\alpha \beta } g^{\gamma \lambda } R \nabla_{\lambda }\nabla_{\gamma }R}{(1 -  q^2 R^2)^{3/2}}=0.
\end{align}
Our objective is to find an asymptotically AdS black brane solution in a four-dimensional symmetric spacetime.\\
To achieve this, we consider the following ansatz,
\begin{equation}\label{metric1}
ds^{2} =-f(r)dt^{2} +\frac{dr^{2} }{f(r)} +\frac{r^2}{L^2} (dx^2+dy^2),
\end{equation}
where  $f(r)$ is the metric function that needs to be determined.\\
Focusing on the $tt$ component of Eq. (\ref{EH}), we have,
\begin{align}\label{eom}
	&2 \Lambda  + \frac{1}{q }\arcsin\Big(\frac{q \big( 2 f(r) + 4 r f'(r)
		+ r^2f''(r)\big)}{r^2}\Big) -  \frac{2 f'(r)+ r f''(r)}{ r \Big(1 -  \frac{q^2 \big(2 f(r) + 4 r
			f'(r) + r^2 f''(r)\big)^2}{r^4}\Big)^{1/2}} \nonumber \\&- \frac{4 q^2 f(r) \big( 2 f(r) + 4 r f'(r) + r^2 f''(r)\big) \big( 4 f(r) + 2 r f'(r) - 4 r^2 
	f''(r)- r^3 f'''(r)\bigr)}{r^6 \Big(1 - \frac{q^2 \bigl( 2 f(r) + 4 r f'(r) + r^2f''(r)\big)^2}{r^4}\Bigr)^{3/2}} \nonumber \\&+\frac{q^2 f'(r)\big( 2 f(r) + 4 r f'(r)+ r^2f''(r)\big) \big(4 f(r) + 2 r f'(r)- 4 r^2 f''(r) - r^3 f'''(r)\big)}{r^5 \Bigl(1 -  \frac{q^2 \big(2 f(r) + 4 r f'(r)+ r^2 f''(r)\bigr)^2}{r^4}\Bigr)^{3/2}}\nonumber \\&+\frac{6q^4 f(r)\big( 2 f(r) + 4 r f'(r)+ r^2f''(r)\big)^2 \big(4 f(r) + 2 r f'(r)- 4 r^2 f''(r) - r^3 f'''(r)\big)^2}{r^{10} \Bigl(1 -  \frac{q^2 \big(2 f(r) + 4 r f'(r)+ r^2 f''(r)\bigr)^2}{r^4}\Bigr)^{5/2}}=0.
\end{align}
Since $f(r)$ cannot be determined exactly, we will approximate it up to the second order in 
$q$.\\
 Following this, we will explore the following forms for $f(r)$ up to the second order of 
$q$,
\begin{equation}\label{f}
	f(r)=f_0(r)+q f_1(r)+q^2 f_2(r),
\end{equation}
Eq. (\ref{eom}) up to first order of $q$ is as follows,
\begin{equation}
r f_0'(r)+f_0(r)+\Lambda  r^2=0.
\end{equation}
The function $f_0(r)$ can be easily determined as,\\
\begin{equation}\label{f0}
	f_0(r)= \frac{C_1}{r}-\frac{\Lambda r^2}{3}.
\end{equation}
Here, $C_1$ represents an integration constant that is associated with the mass of the black hole, specifically $C_1=-2m_0$.\\
 By imposing the condition $f_0(r_h)=0$,, we can solve for 
$f_0(r_h)=0$, as follows,
\begin{equation}
	C_1=\frac{\Lambda r_h^3 }{3},
\end{equation}
Substituting this value of $C_1$ into Eq. (\ref{f0}), we obtain,
\begin{equation}
	f_0(r)= \frac{\Lambda}{3r} (r_h^3-r^3).
\end{equation}
When we consider Eq. (\ref{eom}) at the first order of $q$, it simplifies to,
\begin{equation}
f_1(r) \left( f_0'(r)+\Lambda  r\right)- f_0 (r)f_1'(r)=0.
\end{equation}
The function $f_1(r)$ takes the form,
\begin{equation}
f_1(r)= C_2 e^{\int _{r_h}^{r}\frac{\Lambda  u+f_0'(u)}{f_0(u)}du}=C_2(\frac{1}{r}-\frac{1}{r_h}).
\end{equation}
Next, we can express Eq. (\ref{eom}) at the second order of $q$ as follows,
\begin{align}
&-6 r^5 f_0(r) \left(f_2(r) \left(f_0'(r)+\Lambda  r\right)+f_1(r) f_1'(r)\right)+6 r^5 f_1(r)^2 \left(f_0'(r)+\Lambda  r\right)\nonumber\\&+r^3 f_0(r)^2 \left(6 r^2
f_2'(r)-\left(4 f_0'(r)+r f_0''(r)\right) \left(-r f_0'(r) \left(7 f_0''(r)+3 r f_0^{(3)}(r)\right)+10 f_0'(r)^2+r^2 f_0''(r)^2\right)\right)\nonumber\\&+12 r f_0(r)^4
\left(2 r \left(2 f_0''(r)+r f_0^{(3)}(r)\right)-21 f_0'(r)\right)\nonumber\\&+3 r^2 f_0(r)^3 \left(2 r f_0'(r) \left(28 f_0''(r)+9 r f_0^{(3)}(r)\right)-52 f_0'(r)^2+r^2
f_0''(r) \left(15 f_0''(r)+4 r f_0^{(3)}(r)\right)\right)\nonumber\\&-92 f_0(r)^5=0,
\end{align}
where $C_2=1$.\\
$f_2(r)$ is as follows,
\begin{align}
f_2(r)=&e^{\int^{r}\frac{\Lambda  u+f'_0(u)}{f_0(u)}du}\bigg(C_3+\int^{r}\frac{A(u_2)du_2}{6 u_2^5 f_0(u_2)^2}e^{-\int^{u_2}\frac{\Lambda  u_1+f'_0(u_1)}{f_0(u_1)}du_1}\bigg),
\end{align}
where,
\begin{align}
A(u_2)=&-6 u_2^5 f_1(u_2)^2 \left(f_0'(u_2)+\Lambda  u_2\right)\nonumber\\&-12 u_2 f_0(u_2)^4 \left(2 u_2 \left(2 f_0''(u_2)+u_2 f_0^{(3)}(u_2)\right)-21 f_0'(u_2)\right)\nonumber\\&-6 u_2^3 f_0(u_2)^3 f_0'(u_2) \left(28 f_0''(u_2)+9 u_2 f_0^{(3)}(u_2)\right)\nonumber\\&-3 u_2^2 f_0(u_2)^3 \left(-52 f_0'(u_2)^2+u_2^2 f_0''(u_2) \left(15 f_0''(u_2)+4 u_2 f_0^{(3)}(u_2)\right)\right)\nonumber\\&+u_2^3 f_0(u_2)^2 \left(4
f_0'(u_2)+u_2 f_0''(u_2)\right) \left(-u_2 f_0'(u_2) \left(7 f_0''(u_2)+3 u_2 f_0^{(3)}(u_2)\right)+10 f_0'(u_2)^2+u_2^2 f_0''(u_2)^2\right)\nonumber\\&+6 u_2^5 f_0(u_2) f_1(u_2)
f_1'(u_2)+92 f_0(u_2)^5,
\end{align}
\begin{align}
	f_2(r)&=\frac{r_h \left(160 \Lambda ^4 r_h^8 r^2-135 r_h^3 r+270 r_h r^3-540 r^4+108 r_h^4\right)-360 \sqrt{3} r^5 \tan ^{-1}\left(\frac{2 r+r_h}{\sqrt{3} r_h}\right)}{3645 \Lambda 
		r_h^{10} r^6}\nonumber\\& +\frac{C_3}{r}.
	\end{align}
The event horizon is defined at the point where $f(r_h)=0$, and the constants $C_1$ , $C_2$ and $C_3$ will be determined based on this condition. Since we have $f(r_h)=0$, it follows that\\
\begin{align}
C_3= \frac{r_h \left(160 \Lambda ^4 r_h^{10} -135 r_h^4 +270 r_h^4-540 r_h^4+108 r_h^4\right)-120 \pi \sqrt{3} r_h^5 }{3645 \Lambda 
	r_h^{15}}.
\end{align}
The Hawking temperature for this black hole is given by,
  \begin{align}\label{Temp}
  T &=\frac{f'(r_h)}{4 \pi}=\frac{f_0'(r_h)}{4 \pi}+q\frac{f_1'(r_h)}{4 \pi}+q^2\frac{f_2'(r_h)}{4 \pi}\nonumber\\&=T_0+q T_1+q^2 T_2,
  \end{align}
where,
\begin{align}
	T_0 &=\frac{f_0'(r_h)}{4 \pi}=\frac{1}{4 \pi r_h}+\frac{\Lambda r_h}{4 \pi},\\
	T_1 &=\frac{f_1'(r_h)}{4 \pi}=-\frac{1}{4 \pi r_h^2},\\
	T_2 &=\frac{f_2'(r_h)}{4 \pi}=\frac{r_h \left(320 \Lambda ^4 r_h^9-1485 r_h^3\right)-600 \sqrt{3} \pi  r_h^4-180 r_h^4}{14580 \pi  \Lambda  r_h^{16}}\nonumber\\&-\frac{r_h \left(160 \Lambda ^4 r_h^{10}-297 r_h^4\right)-120 \sqrt{3} \pi  r_h^5}{2430 \pi  \Lambda  r_h^{17}}
\end{align}
The entropy can be found by using Hawking-Bekenstein formula,
\begin{eqnarray}
	A&=&\int d^{2} x \sqrt{-g} |_{r=r_0,t=cte}= \frac{r_{h}^{2} V_{2}}{l^{2}},\nonumber\\
	S&=&\frac{A}{4G} =\frac{r_{h}^{2} V_{2} }{4l^{2}G}, \nonumber\\
	s&=&\frac{4\pi r_{h}^{2}}{l^{2}},   
\end{eqnarray}
where $V_{2}$ is the volume of the constant $t$ and $r$ hyper-surface with radius $r_{+}$ and in the last line we used $\frac{1}{16\pi G} =1$ so $\frac{1}{4G} =4\pi$.
\section{The ratio of shear viscosity to entropy density}
\label{sec3}
\indent The AdS/CFT duality leads to a fluid-gravity duality in the long-wavelength limit. The hydrodynamic equations can be derived from Noether's theorem. In our model, the conservation of the energy-momentum tensor is the equation of motion as follows,
\begin{align}
	& \nabla _{\mu } T^{\mu \nu} =0,\\
	& T^{\mu \nu } =(\rho +p)u^{\mu } u^{\nu } +pg^{\mu \nu } -\sigma ^{\mu \nu },\nonumber\\
	&\sigma ^{\mu \nu } = {P^{\mu \alpha } P^{\nu \beta } } [\eta(\nabla _{\alpha } u_{\beta } +\nabla _{\beta } u_{\alpha })+ (\zeta-\frac{2}{3}\eta) g_{\alpha \beta } \nabla .u]\nonumber\\& P^{\mu \nu }=g^{\mu \nu}+u^{\mu}u^{\nu}, \nonumber
\end{align}
\indent where $\eta$, $\zeta $, $\sigma ^{\mu \nu }$ and $P^{\mu \nu }$ are shear viscosity, bulk viscosity, shear tensor and projection operator, respectively \cite{Kovtun2012, Bhattacharyya,Rangamani}.\\
We want to calculate the shear viscosity using the Green-Kubo formula, which is derived from linear response theory \cite{Ref11,Son:2008zz,Ref13} and is related to the 2-point function of the energy-momentum tensor:
\begin{equation}
	\eta =\mathop{\lim }\limits_{\omega \to 0} \frac{1}{2\omega } \int dt\,  d\vec{x}\, e^{i\omega t} \left\langle [T_{y}^{x} (x),T_{y}^{x} (0)]\right\rangle =-\mathop{\lim }\limits_{\omega \, \to \, 0} \frac{1}{\omega } \Im G_{y\, \, y}^{x\, \, x} (\omega ,\vec{0}).
\end{equation}
We aim to compute the ratio of shear viscosity to entropy density using the formula presented in \cite{Hartnoll:2016tri},
\begin{equation}
	\frac{\eta}{s}=\frac{1}{4 \pi} \phi(r_h)^2.
\end{equation}
We introduce a perturbation in the metric as $\tilde{g}_{\mu \nu} \to g_{\mu \nu} +h_{x\,y}$ where $h_{x\,y}=\frac{r^2}{l^2}\phi(r)$.\\
Next, we substitute $\tilde{g}_{\mu \nu} \to g_{\mu \nu} +h_{x\,y}$ into the action given by Eq.(\ref{action}) and expand the action up to second order in $\phi$,
\begin{align}\label{action2}
	\tilde{S}_2&=\frac{1}{16\pi G}\int d^{4}x  \frac{1}{6 L^2 r^4}\bigg(3 r^2 f(r) \left(r^4 \phi '(r)^2-2 \left(4 f'(r)+r f''(r)\right)^2\right)\nonumber\\&-12 r f(r)^2 \left(4 f'(r)+r f''(r)\right)-r^3 \left(4 f'(r)+r f''(r)\right)^3-8 f(r)^3\bigg).
\end{align}
We then derive the equation of motion for $\phi(r )$ by varying the resulting action with respect to $\phi(r )$,
\begin{eqnarray}
	\left(r f'(r)+2 f(r)\right) \phi '(r)+r f(r) \phi ''(r)=0.
\end{eqnarray}
To solve this equation, we employ Eq. (\ref{f}) and express $\phi(r)$ as follows,
  \begin{equation}
\phi(r)=\phi_0(r)+q \phi_1(r).
 \end{equation}
Considering only the zeroth order in $q$, we find,
\begin{equation}
	\left(r f_0'(r)+2 f_0(r)\right) \phi_0'(r)+r f_0(r) \phi_0''(r)=0.
\end{equation}
The solution for $\phi_0(r)$ is given by,
\begin{align}\label{phi}
	\phi_0(r)=C_5 + C_6\int^{r}\frac{1}{f_0(u)}du.
\end{align}
The behavior of $\phi_0(r)$ near the event horizon is expressed as,
\begin{align}
	\phi_0(r)=C_5+\frac{C_6}{f_0'(r_h)}\log(r-r_h)
\end{align}
To ensure that $\phi_0(r)$ is regular at the event horizon, we apply the regularity condition, yielding,
\begin{eqnarray}
	C_6=0
\end{eqnarray}
For normalization, we set $C_5=1$, leading to the solution,
\begin{align}
	\phi_0(r)=1.
\end{align}
The equation of motion of $\phi$ up to first order of $q$ is as follows,
\begin{align}
&\left(r f_0(r)+2 f_0(r)\right) \phi_1'(r)+r \left(f_0(r) \phi_1''(r)+f_1(r) \phi_0''(r)\right)\nonumber\\&+\left(r f_1'(r)+2 f_1(r)\right) \phi_0'(r)=0
\end{align}
This equation is complex to solve. Thus, we approach a solution near the event horizon by substituting $f_0(r_h)=f_1(r_h)=0$. Consequently, the solution for $\phi_1(r)$ takes the form,
\begin{equation}\label{phi-1}
	\phi_1(r)=C_7+ \int^{r}\frac{C_8-f_1(u) u^2 \phi_0'(u)}{f_0(u) u^2}du.
\end{equation}
By substituting $\phi_0'=0$ in Eq. (\ref{phi-1}) we have,
\begin{equation}
	\phi_1(r)=C_7+\frac{C_8}{4 \pi T_0 r_h^2}\log (r-r_h).
\end{equation}
To be regular at the horizon we should get rid of $\log (r-r_h)$ by choosing,\\
\begin{equation}
	C_8=0,
\end{equation}
\begin{equation}
	\phi_1(r)=C_7.
\end{equation}
If we repeat this process for $\phi_2$, then we will have something similar to $\phi_1$.
\begin{equation}
	\phi_2(r)=C_{9}.
\end{equation}
The expression for $\phi(r)$ is given by,
\begin{align}
	\phi(r)&=1+q \phi_1(r)+q^2 \phi_2(r)=1+C_7 q+C_9 q^2.
\end{align}
Since the function $\phi$ is a wavefunction and, due to the concept of probability, it must be less than or equal to one, we consider the constant $C_7$ and $C_9$ to be negative.\\ 
\begin{align}
	\phi(r)&=1+q \phi_1(r)+q^2 \phi_2(r)=1-|C_7| q-|C_9| q^2.
\end{align}
The ratio of shear viscosity to entropy density, up to second order in $q$, can be expressed as,\\
\begin{equation}
	\frac{\eta}{s}=\frac{1}{4 \pi} \phi(r_h)^2=\frac{1}{4 \pi} (1-|C_7| q-|C_9| q^2)^2=\frac{1}{4 \pi} \big(1-2 q |C_7|-2q^2 |C_9|^2\big).
\end{equation}
For our model, the KSS bound is not satisfied at second order in $q$.\\
 The Kovtun-Son-Starinets (KSS) bound \cite{Policastro2002} asserts that $\frac{\eta}{s} \geq \frac{1}{4 \pi} $ for all quantum field theories.\\
 In the case of Einstein-Hilbert gravity and its field theory dual, this bound is saturated. However, this saturation is not maintained in higher derivative gravities \cite{Sadeghi:2015vaa, Sadeghi:2022kgi,Iqbal:2008by}, massive gravity \cite{Hartnoll:2016tri}, \cite{Sadeghi:2015vaa}, \cite{Parvizi:2017boc},\cite{Sadeghi:2018ylh}, \cite{Alberte:2016xja}, and deformed black branes \cite{Ferreira-Martins:2019svk}.\\
\section{Conclusion}

\noindent We have presented the arcsin AdS black brane solution and calculated the ratio of shear viscosity to entropy density for this model. While there is a conjecture that this ratio is saturated in Einstein-Hilbert theory, it can be violated in the context of higher derivate gravity. Specifically, the computed ratio of shear viscosity to entropy density for our model does not comply with the Kovtun-Son-Starinets (KSS) bound \cite{Kovtun:2004de} when $\phi_1<0$. This bound indicates that the value for Einstein-Hilbert gravity is given by 
$\frac{\eta }{s}=\frac{1 }{4 \pi}$. Additionally, we find that $\frac{\eta }{s}$ is inversely proportional to the square of the coupling constant in the field theory, specifically $\frac{\eta }{s} \sim \frac{1 }{\lambda^2}$. This suggests that the coupling of the field theory dual to our model differs from that of the Einstein-Hilbert gravity dual.\\

\noindent {\large {\bf Acknowledgment} } We are grateful to the anonymous referee of the International Journal of Modern Physics A (IJMPA) for the valuable feedback and insightful comments, which have helped us significantly improve the quality and clarity of this manuscript.

\vspace{1cm}
\noindent {\large {\bf Data Availability } } Data generated or analyzed during this study are provided in full within the published article.

\end{document}